\documentclass[12pt,preprint]{aastex}

\usepackage{natbib}

\newcommand{\be}{\begin{equation}}
\newcommand{\ee}{\end{equation}}
\newcommand{\nn}{\mbox{} \nonumber \\ \mbox{} }
\newcommand{\ba}{\begin{eqnarray}}
\newcommand{\ea}{\end{eqnarray}}
\newcommand{\om}{\omega}
\newcommand{\Alfven}{Alfv\'{e}n }

\newcommand{\B}{{\bf B}}

\newcommand\etal{\textit{et al.\ }}
\newcommand\eg{\textit{e.g.,\ }}

\newcommand{\Bf}{{magnetic field}}
\newcommand{\Bfs}{{magnetic fields}}

\newcommand{\NS}{neutron star}
\newcommand{\NSs}{{neutron stars}}

\newcommand{\cm} {\rm cm}
\newcommand{\ergs} {\rm erg\,  s^{-1}}
\newcommand{\cms} {\rm cm\,  s^{-1}}

\begin{document}

\title{Super-Eddington accretion in ultra-luminous  neutron star binary}

\author{Maxim Lyutikov}
\affil{
Department of Physics and Astronomy, Purdue University, 
 525 Northwestern Avenue,
West Lafayette, IN
47907-2036}

\begin{abstract} 
We discuss  properties of the 
ultra-luminous $X$-ray source in the galaxy M82, NuSTAR J095551+6940.8, containing an accreting \NS.
The neutron star has  surface  magnetic field $ B_{NS} \approx 1.4 \times 10^{13 } \,  {\rm G}$ and experiences accretion rate of $9 \times 10^{-7} M_\odot {\rm  \, yr}^{-1} $. The magnetospheric radius, close to the corotation radius, is $\sim 2 \times 10^8$ cm.
The accretion  torque on the neutron star is reduce well below  what is expected in a simple magnetospheric  accretion  due to effective penetration of the stellar magnetic field into the disk beyond the corotation radius.  As a result, the radiative force of the surface emission  does not lead to  strong coronal  wind, but pushes plasma along magnetic field lines towards the equatorial disk. The neutron star is nearly an orthogonal rotator, with the angle between the rotation axis and the magnetic moment  $\geq 80$ degrees.
Accretion occurs through optically thick -- geometrically thin and  flat accretion ``curtain'',  which cuts across  the polar cap. High radiation pressure  from the neutron star surface is nevertheless smaller than that the ram pressure of the accreting material flowing through the  curtain, and thus fails to stop the accretion. 
At distances below few stellar radii the  magnetic suppression of the scattering cross-section becomes important.
The  $X$-ray luminosity (pulsed and  persistent components) comes both from the neutron star surface as a hard $X$-ray component and as a soft component from  reprocessing  by the accretion disk.

 \end{abstract}

\section{Introduction}

Ultra-luminous $X$-ray sources (ULXs) is a class of objects observed in nearby galaxies with luminosities in the $10^{39}-10^{41} \ergs$ range, not associated with the central black hole
\cite{2006csxs.book..475F,2007Ap&SS.311..203R,2011NewAR..55..166F}. Such luminosities exceed Eddington luminosity of a stellar-mass central object. Until recently, the discussion on the nature of these object concentrated on a choice between jetted/beamed  emission  from stellar mass black holes or accretion onto intermediate, $M \sim 100-1000 M_\odot$, black hole (possibly also producing a jet)
\citep{2003Sci...299..365K,2010MNRAS.408..234M,2012MNRAS.423.1154S,2013MNRAS.435.1758S}. 

To a great surprise,  \cite{2014Natur.514..202B} discovered 1.37 second pulsations and 2.5 day orbital modulations in  a ULX located in M82, NuSTAR J095551+6940.8, - a clear signature of an accretion on a rotating \NS.  The companion has a minimal mass of $5.2 M_\odot$, classifying the system as a high mass $X$-ray binary. This implies isotropic  luminosity of the order of a hundred of Eddington luminosities.  \citep[][discussed a possibility that ULXs are  young, rotationally powered pulsars; slow spin and  detection of the companion in NuSTAR J095551+6940.8 are evidence of accretion- not rotaion-powered source.]{2013MNRAS.431.2690M}


Super-Eddington accretion cannot proceed isotropically (Sec. \ref{1}). Slightly super-Eddington accretion rates \citep{1998ApJ...494L.193S,2011MNRAS.413.1623D}  are possible due to instabilities in the accretion flow. In most cases considered so far, super-Eddington accretion rates were discussed for accretion onto black holes, not \NSs\ \citep[see, though][]{1992ApJ...388..561A}. Applicability  of these models to the case of \NS\ deserves  separate investigations. Here we discuss a  possible scenario of a geometrically thin, but highly optically thick accretion  disk that  brings the matter on the surface of a \NS\ while producing highly super-Eddington luminosity outside of the disk. 

\section{Accretion properties}
\label{1}

Let us  first apply the standard model of accretion in binary systems \citep[\eg][]{1972A&A....21....1P,2002apa..book.....F} to the observations of NuSTAR J095551+6940.8. As we will see, the standard accretion  model will fail to explain the system's properties, but a simple modified theory succeeds. (In what follows we assume a \NS\ with a mass of $1.4 M_\odot$.)

The system's luminosity is produced due to the accretion of matter on the surface of a \NS. The $X$-ray luminosity $L_X$ then is a fraction of the accretion power  \citep{1976MNRAS.175..395B} 
\ba &&
L_{acc} = \dot{M} GM /R_{NS}
\nn &&
L_X= \eta L_{acc}, \, \eta \sim 0.5
\ea
(some of the accreted energy is radiated in neutrinos and some is conducted into the \NS).
Parametrizing the luminosity to the Eddington luminosity, $l_E = L_X/L_E$, $L_E = 4 \pi c G M_{NS} /\sigma_T$, and the  mass accretion rate to the Eddington rate, $\dot{M}_E = L_E/c^2$,
\ba &&
m_E = \dot{M}/M_E
\nn &&
r_G =  2 GM/c^2,
\ea
the required $m_E$  is 
\be
m_E = 2 {l_E \over \eta} {R_{NS} \over r_G} \approx 300 \left( {l_E \over 30} \right)\, \left( {\eta \over 0.5 } \right), 
\label{mE}
\ee
(in physical unites this corresponds to the accretion rate of $5.6 \times 10^{19} {\rm g \, s}^{-1} = 8.5 \times 10^{-7} M_\odot {\rm  \, yr}^{-1} $). Though the observed luminosity is close to $l_E \sim 100$, the emission  is anisotropic; hence we parametrize it  to the value of  $30 L_{Edd}$. 

For a given accretion rate $m_E$ the optical depth to Thompson scattering is
\be
\tau_T = n \sigma_T r =  {m_E} \sqrt{r_G \over 2 r}
\ee
Thus, isotropic flow would  become optically thick at 
\be 
r \sim 2 m_E^2 r_G \approx  7 \times 10^{10} \cm
\ee
This would preclude observations of the rotational modulations from the \NS\ -  the accretion must be anisotropic. 

For a given accretion rate the (magnetospheric)  \Alfven radius is located at
\be
r_A  = \left( { B_{NS}^4 R_{NS}^{12} \over 2 G M \dot{M}^2} \right)^{1/7}. 
\label{rm}
\ee
(anisotropy of the accretion flow will reduce $r_A$, but not substantially due to steep dependence of dipolar magnetic pressure on radius.)
It is expected that a system quickly approaches equilibrium, when the \Alfven radius equals the corotation radius $r_c$
\be
r_c = (G M /\Omega^2)^{1/3} = 2 \times 10^8 \cm
\label{rc}
\ee
The requires  surface \Bf\ is 
\be
B_{NS} =1.4 \times 10^{13}\, \left( {l_E \over 30} \right)^{1/2}\, \left( {\eta \over 0.5 } \right)^{- 1/2} \, {\rm G}
\label{BNS}
\ee
This value is somewhat larger than the conventional {\NSs}' \Bf\  of  $10^{12}$ G, but still within  the range of population of young pulsars; this is also smaller that the magnetars' \Bfs\ \citep{TD95}. 
 
All of the above relations follow from the standard accretion models \citep[\eg][]{2002apa..book.....F}. On the other hand,  a simple torque prescription for the accreting matter is inconsistent  with the above, as already pointed out by \cite{2014Natur.514..202B}. The accretion torque due to the material accreting from the disk is
\be
I \dot{\Omega}=T_d= \dot{M} \sqrt{ GM r_d} ,
\label{torque}
\ee
where $r_d$ is the distance at which torque is applied (usually the inner edge of the disk located at $r_A$). Together with the measured spin-up rate $\dot{P}= - 2 \times 10^{-10}$ this 
implies
\be 
r_d m_E^2 = 5 \times 10^{10}\cm
\label{rd}
\ee
For $m_E \sim 300$ this would give  for $r_d$ an extremely small value, less than the \NS\ radius. Thus the accretion torque is more than 10 times smaller than expected from  simple accretion models.

Which of the above model  assumptions could have gone wrong? The first suspect is the assumption of the corotation at the inner edge of the disk, $r_A = r_c$, \eg\ due to the  intermittent  accretion rate  \citep{2014Natur.514..202B}. We disfavor this possibility, since, first, 
with a period of $1.37$ seconds and the spin-up rate of $\dot{P}= - 2 \times 10^{-10}$, the time to reach equilibrium, where the inner edge of the disk coincides with corotation, is very short, about 200 hundred years. It is then unlikely that we are observing the source at a special time, with substantially different $\dot{M}$. Second, from Eqs. (\ref{rc},\ref{rd}) it follows that the instantaneous accretion rate should be $<m_E> \approx 16$, which is nearly 20 times lower than the instantaneous accretion rate (\ref{mE}). Higher accretion efficiency,  as well as more  anisotropic emission from the accretion column may further alleviate the problem somewhat; still the instantaneous accretion rate must be more than an order of magnitude higher than the average one.

As a resolution of the contradiction between the required high accretion  rate and relatively small spin-up rate, we suggest that the torque  on the star is reduced with respect to pure disk torque $T_d$, Eq. (\ref{torque}), due to the penetration of the \Bf\ from the star into the accretion disk  beyond the corotation, as discussed by \cite{1973ApJ...184..271L,1976ApJ...207..914A,1979ApJ...232..259G,1979ApJ...234..296G,1987MNRAS.229..405C,1990A&A...227..473A,1996A&A...307.1009B,2000MNRAS.317..273A,2004ApJ...606..436R,2007ApJ...671.1990K,2013MNRAS.435.2633N,1987A&A...183..257W}. Magnetic field lines from the central star penetrate the disk beyond corotation, where they are dragged by the accreting flow with the angular velocity smaller than the angular velocity of the star, thus removing the angular momentum from the star. The additional angular momentum thus deposited in the disk is removed through the conventional disk turbulence. 

 The exact value of the magnetic torque depends on how the \Bf\ of the star penetrates the disk and how it  "slides" through the accretion disk - both depend on the assumed prescription for turbulence-generated disk resistivity \citep{2000MNRAS.317..273A}.
 Assuming effective penetration of the \Bf\ into the accretion disk, so that on the disk the toroidal \Bf\ is  $B_\phi \approx  - B_z ( 1- \om_K/\Omega_{NS}) $ with the poloidal component  approximately given by the dipolar field, 
   the magnetic torque is 
\be
T_M
\approx - C (B_{NS}^2 R_{NS}^6) / r_A^3 ,
\ee where $C$ is some diminutional constant \citep[\eg][]{2004ApJ...606..436R}. Since our  estimate of the expected torque is nearly two  orders of magnitude larger than given by Eq. (\ref{torque}), the matter  torque is nearly compensated by the magnetic torque. Equating $T_d=T_M$ gives the magnetospheric radius
\be
r_t = C^{2/7} r_A
\ee
Thus,   the parameter $C$   should be of the order of unity (though it enters with a small power) to accommodate  large accretion rate, large corotation radius and yet a small torque. 
For the toroidal \Bf\ scaling mentioned above, $C=1/21$, $C^{2/7} = 0.42$, consistent with the requirement that the magnetospheric  radius should be somewhat smaller than the corotation radius for accretion to occur. 

Dragging of \Bfs\ lines through the disk due to turbulent resistivity (as opposed to twisting due to  the frozen-in condition) is valid as long as the difference between the relative linear velocity of solid body rotation of the field lines and of the Keplerian motion in the disk is smaller than the sound speed in the disk, 
$ (\Omega  - \sqrt{ GM_{NS}/r^3}) r \leq  c_s $. For a standard $\alpha$-disk  \citep{ShakuraSunyaev,2002apa..book.....F} this is satisfied  only in a fairly narrow region near $r_A$ -  the sound speed at $r_A$ is fairly small, $c_s \approx 10^7 \cms$. Thus, the fastness parameter \citep{1979ApJ...234..296G} must be close to unity. 
(N.B.: the light cylinder radius $r_{LC} = 6 \times 10^9$ cm is only 30 times larger than the \Alfven radius).

       \section{Accretion disk and the curtain}

 \subsection{Disk thickness}

Using the standard accretion disk theory \citep{ShakuraSunyaev,2002apa..book.....F} and the parameters estimated above,  we can find the thickness of the accretion disk at the \Alfven radius, 
\be 
H \approx 6 \times 10^6 \alpha^{-1/10} \cm,
\ee
where $\alpha$ is the disk viscosity parameter \citep{ShakuraSunyaev}.
The disk is slightly radiatively dominated, with the ratio of radiation pressure to particle pressure $\sim 3$.  Thus, the accretion disk is thin, $\Delta \equiv H/r \sim 0.03$. The  internal  temperature at the inner edge is $T \sim 3 \times 10^6 $K $= 0.3 $ keV. 

In what follows we assume that the relative disk  thickness $\Delta$ at the \Alfven radius (and not the radial extent of the \Bf\ penetration into the disk)  determines the relative thickness of  the accretion curtain. As a consequence, at a radius $r< r_A$ the cross-section of the accretion curtain is $S= \Delta r r_A$.

 \subsection{Effects of radiative pressure in the corona}

For super-Eddington accretion to occur, most of the infalling material should be shielded from the radiation. Next, we discuss possible geometry of the accretion flow that may allow such process to occur. 

\begin{figure}[h!]
\includegraphics[width=.99\linewidth]{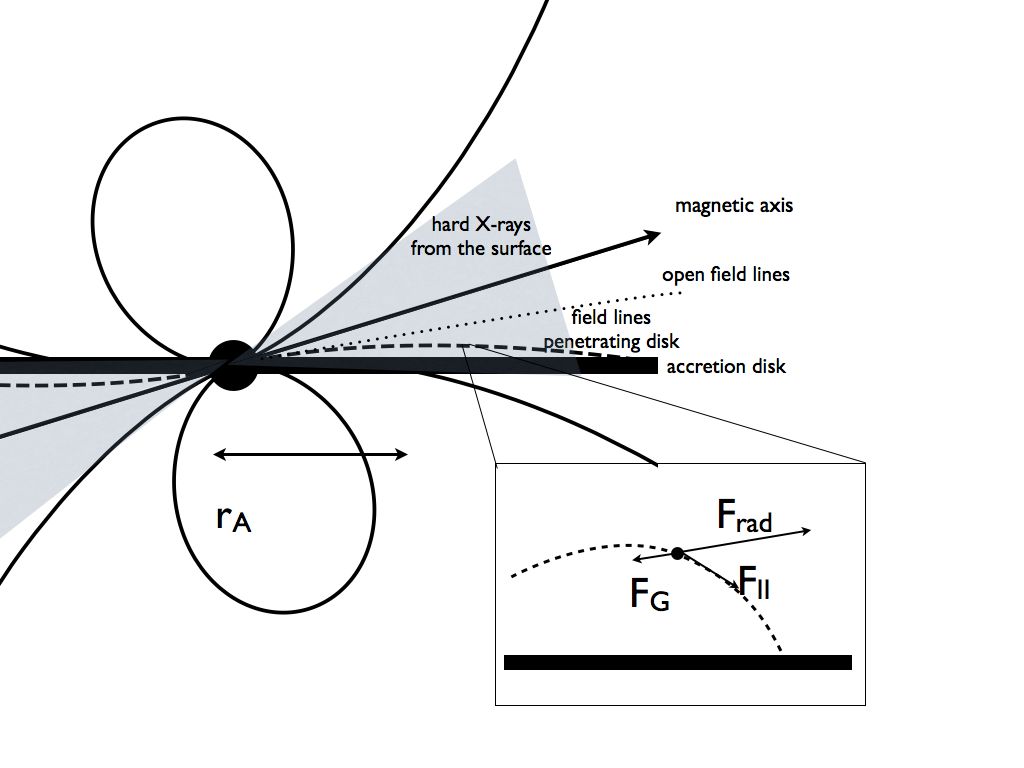}
\caption{Geometry of the accretion curtain. The pulsar is nearly orthogonal rotator with inclination $\geq 80^\circ$ (rotation axis is vertical, perpendicular to the disk surface). Accretion disk is stopped at the \Alfven radius $r_A$ by the magnetospheric \Bf, and  accretes along highly inclined field lines close to the magnetic pole.  Accretion occurs in a narrow accretion curtain that cuts through the polar cap. Strong radiative force $F_{\rm rad}$ of the surface hard $X$-ray emission is higher than the gravity force $F_G$; the resulting force  parallel to the \Bf\ $F_\parallel$   clears most of the accretion column and pushes  the particles along \Bfs\ lines toward the accreting curtain (see  the insert). Magnetic fields penetrate the disk beyond the \Alfven radius (dashed field lines);  at larger distances the stellar \Bf\ lines remain open (dotted line).}
\label{picture-ULX}
\end{figure}

Presence of the \Bf\ penetrating the disk is the key. 
Large radiative flux exerts a radial force on the falling matter, mostly on electrons. In  \Bf\ the    radiation pressure leads, first,   to   drift of electrons in a direction perpendicular both to the force and the local \Bf, and, second,  a force parallel to the \Bf\  that in the accretion geometry  {\it pushes the accreting matter towards the magnetic equator}, see Fig. \ref{picture-ULX}. 
Thus, high  central luminosities, combined with penetration of the disk by the stellar \Bf\ confine the accreting matter to  a narrow accretion curtain.

If a force ${\bf F}$ acts on a particle in \Bf, the particle experiences a drift with velocity
\be
{\bf v}_d = {{\bf F} \times \B \over e B^2}. 
\ee
For an electron in dipolar \Bf\ acted upon by the radiative force from a source of luminosity $L$ at the origin this drift velocity equals
\be
v_d  \approx  { \sigma _T L r \over e  B_{NS} R_{NS}^3}
\ee 
(at the \Alfven radius (\ref{rm}) this evaluates to several centimeters  per second.) The corresponding velocity of the ions is smaller due to smaller scattering cross-section. This  drifts of electrons produces a  current that modifies the \Bf, but  since we do not expect substantial amount of matter directly exposed to radiation (most is screened), such a modification of the \Bf\ is expected to be small.

The kinetic pressure at the edge of the disk, $P \sim 10^{12} \alpha^{-9/10}$ (in cgs units)  \citep{ShakuraSunyaev,2002apa..book.....F}  slightly  dominates over the radiation pressure, $P_{\rm rad}  \sim L/(4 \pi r_A^2 c) \sim 3  \times 10^{11}$.   Importantly, due to effective penetration of the stellar  \Bf\ into the disk this radiative force does not lead to the formation of the wind.
For super-Eddington luminosities the parallel repulsive force on the electron is larger than the parallel component of the attractive gravitational force on the ions. Electrons and ions are strongly coupled electrostatically. Both electrons and ions move away from the star, but their motion is restricted by magnetic field. As a results, plasma is radiatively pushed from the corona towards the disk. Thus, the penetrating \Bf\ leads to effective ``magneto-radiative'' compression of the disk.
We conclude that super-Eddington luminosity through the disk corona does not lead to the formation of the coronal wind: it keeps the matter in a disk.

 The concepts of the well-defined magnetospheric radius and of the effective field penetration into the accretion disk are somewhat contradictory. The assumption is that subdominant fields are turbulently mixed into the disk, while dominant \Bfs\ within the \Alfven radius truncate the disk.

(As discussed above, only the parts of the disk close to the inner edge are penetrated by the stellar \Bfs\ and are thus  subject to ``magneto-radiative'' compression. Parts of the disk well beyond the \Alfven radius have open field  lines and might be subject  to radiative oblation. High apparent accretion rates indicate that such ablation is not important, probably due to high incidence angle of the incoming radiation, as well as possible shadowing by the inner edge of the disk).

Note also, that at the magnetospheric radius $r_A$ the radiation pressure, $p_{\rm rad} \approx L/(4 \pi r_A^2 c)$ approximately equals the \Bf\ pressure $B(r_A)^2 /(8\pi)$. Thus, the coronal  field lines can be ``pushed'' into the disk by the radiation by exerting a radial force on  stray particles in the corona.

\subsection{Optically thick accretion curtain}
As discussed above, 
due to high radiation pressure material cannot flow along the \Alfven surface - it is pushed back towards the disk by the radiation pressure (if the stellar \Bf\ did not penetrate the disk, the matter would be blown away). The material can still accrete if   the flow from the accretion disk proceeds  along  a nearly flat, optically thick accretion curtain. For this, the  inclination angle between the magnetic and rotation axis should be large, $\geq 80^\circ$, Fig. \ref{picture-ULX}. In this case accretion proceeds along a nearly flat surface, which cuts across the polar column (see \S \ref{shape}). 
The optical depth across the accretion curtain at radius $r$ is 
\be
\tau_T \approx \Delta  n \sigma_T  r \approx 4 \pi m_E {\sqrt{ G M_{NS} r } \over c r_A} 
\ee
where we estimated the density $n= \dot{M}/( m_p S v)$, cross-section of the column $S = \Delta r r_A$ and infall velocity as a free-fall $v \sim \sqrt{ G M_{NS}/r}$. 
At the \Alfven radius $\tau_T \approx 120 \gg 1$.  

High optical depth across the accretion curtain and the inner disk has several important implications. First, a large fraction of the hard $X$-ray surface emission is absorbed by the accretion disk and is then re-radiated. The expected disk  luminosity
\be
L \approx \pi r_A^2 \sigma_{SB} T^4 \approx 10^{41} T_{\rm keV}^4 \ergs 
\ee
matches the observed luminosity \cite{2014Natur.514..202B}.

Second, high  optical depth across the accretion curtain implies that accreting matter is mostly shielded from the photons and falls directly onto the surface with nearly free-fall velocity (see also \S \ref{lastfew}).
Finally, high  optical depth across the accretion curtain implies that plasma cannot efficiently cool via sideways emission. 

\subsection{The last few radii}
\label{lastfew}

At sufficiently small distance from the star the magnetic suppression of the scattering cross-section may become important for the $X$-mode (polarized in a direction perpendicular to the local \Bf\ and the direction of propagation), $\sigma_X \approx \sigma_T (\om/\om_B)^2$, where $\om$ is the photon frequency. (For nearly radial \Bf\ near the magnetic pole most of the surface emission  propagates nearly along the field, corresponding to X-mode polarization.)  For the given estimate of the \Bf\ (\ref{BNS})
we can find  the distance $r_{Edd,X}$ where the radiative force on electrons $\sim L/(4 \pi r^2 c) \sigma_T (k_B T/ \hbar \om_B)^2$ becomes comparable to the gravitational pull on ions:
\be
r_{Edd,X} =\sqrt{2} \pi^{1/6} { ( e \hbar )^{1/3} ( G M)^{5/18} m_p ^{1/6} R_{NS}^{1/6} \over (\eta \sigma_T)^{1/6} (k_B T)^{1/3} c^{1/6} m_e^{1/3} \Omega^{7/18}}
\approx
1.5 R_{NS} T_{\rm NS,10  keV}^{-1/3}
\ee
Thus, material that gets  to $r< r_{Edd,X} $ is freely accreted onto the \NS.

It is not clear to what temperatures the surface of the \NS\  will be heated in such high accretion rates. The direct heating models \citep[\eg][]{1975JETP...41...52B,1984ApJ...278..369H} assume low accretion rates, $m_E \ll 1$ and smaller \Bfs.  If for $m_E \gg 1$  the effective surface temperature is  tens of keV, then  for such high surface temperatures magnetic suppression of the scattering cross-section is not important at the surface \Bf\ (\ref{BNS}). In addition, resonant scattering may start to become important for surface temperatures $\geq 100 $ keV. Thus, if  the  surface temperatures  reaches  $\geq 50$ keV  the accretion curtain is mostly shielded from the surface emission all the way to the surface. 

On the other hand, for smaller surface temperatures  the accreting layer where  $\tau \sim 1$   experiences a repulsive radiative force (the thickness of the layer within which the optical depth becomes of the order of unity is $\delta r \approx \sqrt{2} r_{Edd,X}^{3/2}/( \sqrt{r_G} m_E) \approx 2 \times 10^{4} $ cm). Can this force stop the flow?  Comparing the radiation pressure $p_{\rm rad} \sim L/(4 \pi r^2c )$ with the ram pressure of the infalling material $p_{\rm ram} =  \rho v^2, \, \rho = \dot{M}/( S v)$, we find that the radiative pressure  is  smaller than the ram pressure,
\be
{ p_{\rm rad} \over p_{\rm ram}} = {\Delta \eta \over 4 \sqrt{2} \pi} \sqrt{r_G \over r} {r_A \over R_{NS}}  \approx 0.1 
\ee
where the numerical estimate is given for $r=R_{NS}$.  Thus, the radiation pressure cannot stop the flow confined to a narrow accretion curtain. 
(For isotropic flow $p_{\rm rad}$ equals $p_{\rm ram}$ at $r/R_{NS} \approx (2/\eta^2) R_{NS}/r_G \approx 20$.)

(If the radiative pressure was high enough to stop the accreting flow, the highly shielded free-falling accretion flow would lead to accumulation of matter at the $\tau \sim 1$ surface. As a result,  the flow would still proceed through formation of narrow accretion channels, somewhat similar to what has been discussed by  \cite{1998ApJ...494L.193S,2011MNRAS.413.1623D}; the photon bubble instability can also be important \citep{1992ApJ...388..561A}.)

It is not clear if the accreting matter should pass through a shock near the surface \citep[\eg][]{2002AJ....123.2019K,2002ApJ...578..420R} or can bombard the surface directly \citep{1984ApJ...278..349B,2008PThPh.119..739K}. If the shock forms, an optically thin accretion column at $r \leq 2 R_{NS}$ might be an important source of high energy emission.

\begin{figure}[h!]
\includegraphics[width=.99\linewidth]{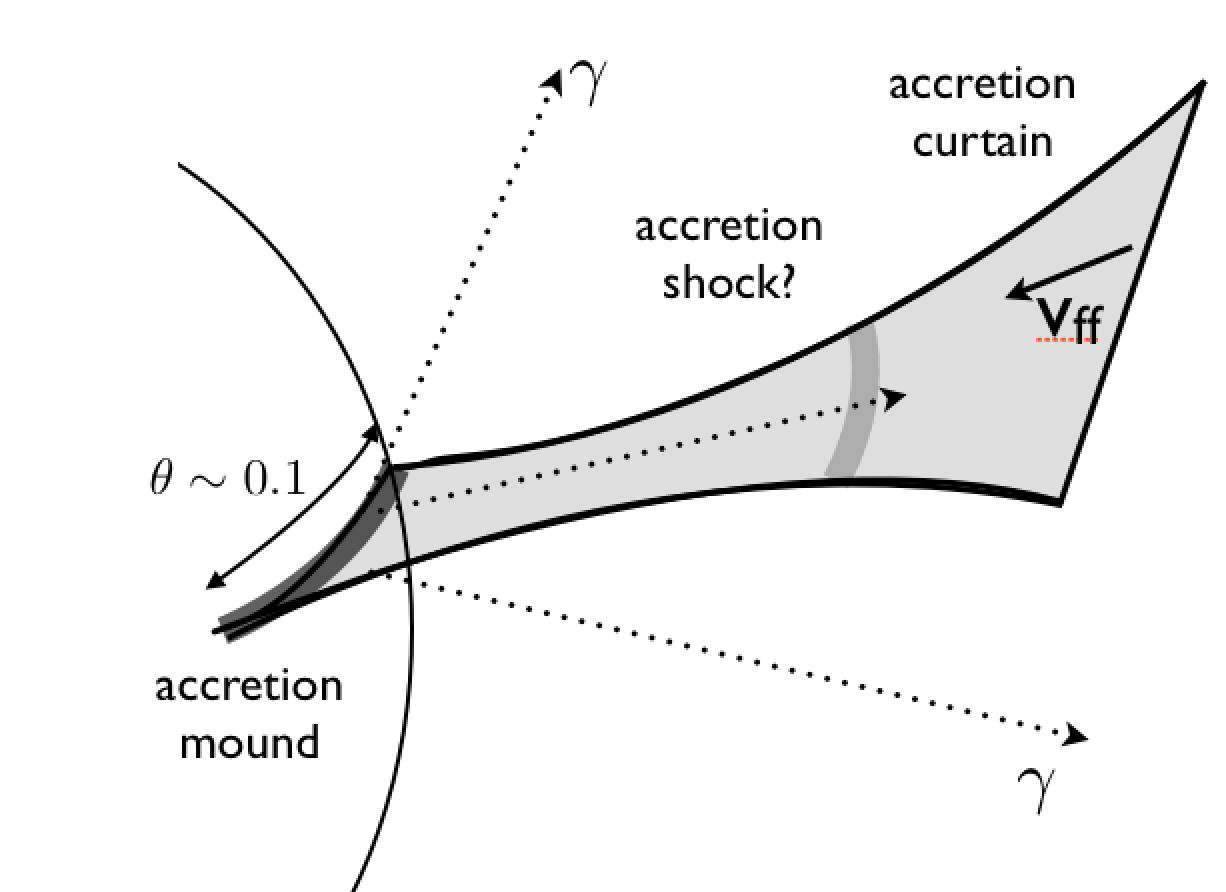}
\caption{Structure of the accretion curtain close to the \NS. The curtain is nearly flat. The flow is in a free-fall. Below $\sim 2 $ \NS\ radii the magnetic  suppression of the scattering cross-section allows material to accrete. An accretion shock may form in the flow, even though at higher radii the radiation pressure is still smaller than the ram pressure of the accreting material. The accretion mound is nearly strait narrow region occupying $\sim 0.1 $ in polar angle. Most of the hard $X$-ray radiation emitted by the mound (dotted lines)  escapes the accretion flow. }
\label{accretioncurtain}
\end{figure}

  \subsection{The shape of the accretion curtain}
  \label{shape}
 
Consider magnetic dipole inclined by angle $\theta_d$ with respect to the accretion disk's normal. Let's assume that the inner  edge of the disk corresponds to a fixed magnetic pressure $p_A$. Also, assume that the disk occupies the plane $\theta = \pi/2$. (It is expected that any corrugation of the disk are smoothed out by the radiation pressure from the central star.) The inner edge of the disk maps  along the dipolar \Bf\ lines on the points on the surface given by 
\ba &&
\sin \theta = {\sqrt{ 1 - \sin ^2 \theta_d \cos^2 \phi} 
\sin \theta_0 \over ( 1+ 3 \sin ^2 \theta_d \cos ^2\phi)^{1/12}}, \, - \pi/2 < \phi < \pi/2
\nn &&
\sin \theta_0= (p_A/B_{NS}^2)^{1/12} = \sqrt{R_{NS}/r_A} \ll 1
\label{curt}
\ea
For aligned case, $\theta_d =0$, Eq. (\ref{curt})  gives a circle $\theta =  \theta_0$, while for orthogonal case,  $\theta_d =\pi/2$,  Eq. (\ref{curt})  gives a line passing though the magnetic pole and expending to $\theta= \theta_0, \, \phi = \pm \pi$. 
In case of nearly orthogonal rotator the accretion will proceed along a nearly flat ``curtain", that extends by $\Delta \theta = \pm \sqrt{R_{NS}/r_A} \sim 0.1$ in the azimuthal direction. For the angle between the rotational and magnetic axes $\theta_d \approx 80 ^\circ$ the curvature of this accretion  curtain near the star  is small,  $\sim (\pi/2 - \theta_d)/ 2^{7/6} \sim 
   0.07$,  Fig. \ref{accretioncurtain}


  \section{Discussion}
  
  In this Letter we discuss the accretion properties of the highly  super-Eddington  source ULX  NuSTAR J095551+6940.8. We suggest that accretion in this source proceeds in a new regime, which was not studied previously - through an optically thick accretion curtain. Previous studies of slightly super-Eddington accretion flows \citep{1976MNRAS.175..395B} concluded  that most of the   accretion energy is radiated through the walls of the accretion column. Highly super-Eddington accretion flows require that most of the accreting matter be shielded from the photons, and, thus,  cannot radiate  efficiently during the infall. 
  
  Several factors are needed for the such super-Eddington  luminosity. First, the externally-supplied accretion rate should be sufficiently high. This most likely requires a Roche lobe overflow with either a He-rich companion or a main sequence H-rich companion (Kalogera \etal\ in prep.). 
   Second, the accretion stage should proceed in a new, previously unexplored regime of optically thick accretion curtain. A needed requirement is that the accreting \NS\ should be a nearly orthogonal rotator (otherwise an accretion flow along the \Alfven surface need to cover a large solid angle is thus will be exposed to large radiation pressure).  
    
   We argue that large accretion  rates combined with  a  small spin-up rate implies that the \Bf\ of the \NS\ effectively penetrates the accretion disk,  removing most of the angular momentum transferred to the star by the falling material.   The suppressions of the torque on the \NS\  due to \Bf\ penetration has  has been previously discussed   by \cite{1979ApJ...234..296G,1987MNRAS.229..405C,2000MNRAS.317..273A,1987A&A...183..257W}. This conclusion  might have important  implications for the estimates of the \Bfs\ in the accreting \NSs.
  For example, the surface  \Bf\  found using the luminosity and the assumption of corotation, Eq. (\ref{BNS}), is nearly four times  larger that  would have ben inferred  using  the spin-up rate  and the assumption of corotation,
  \be
  B'_{NS} = {  \sqrt{2 G M_{NS} I_{NS} \dot{\Omega}}  \over R_{NS}^3 \Omega }= 3 \times 10^{12}\,  {\rm G}
  \ee

  We would like to thank Matteo Bachetti, Victoria Kalogera,  Sergey Komissarov, Timothy Linden, Sergey Popov, Mikhail Revnivtsev, and Nir Shaviv for discussions. 
\bibliographystyle{apj}
  \bibliography{/Users/maxim/Home/Research/BibTex}

\begin{thebibliography}{35}
\expandafter\ifx\csname natexlab\endcsname\relax\def\natexlab#1{#1}\fi

\bibitem[{{Agapitou} \& {Papaloizou}(2000)}]{2000MNRAS.317..273A}
{Agapitou}, V. \& {Papaloizou}, J.~C.~B. 2000, \mnras, 317, 273

\bibitem[{{Aly} \& {Kuijpers}(1990)}]{1990A&A...227..473A}
{Aly}, J.~J. \& {Kuijpers}, J. 1990, \aap, 227, 473

\bibitem[{{Arons}(1992)}]{1992ApJ...388..561A}
{Arons}, J. 1992, \apj, 388, 561

\bibitem[{{Arons} \& {Lea}(1976)}]{1976ApJ...207..914A}
{Arons}, J. \& {Lea}, S.~M. 1976, \apj, 207, 914

\bibitem[{{Bachetti} {et~al.}(2014){Bachetti}, {Harrison}, {Walton},
  {Grefenstette}, {Chakrabarty}, {F{\"u}rst}, {Barret}, {Beloborodov}, {Boggs},
  {Christensen}, {Craig}, {Fabian}, {Hailey}, {Hornschemeier}, {Kaspi},
  {Kulkarni}, {Maccarone}, {Miller}, {Rana}, {Stern}, {Tendulkar}, {Tomsick},
  {Webb}, \& {Zhang}}]{2014Natur.514..202B}
{Bachetti}, M., {Harrison}, F.~A., {Walton}, D.~J., {Grefenstette}, B.~W.,
  {Chakrabarty}, D., {F{\"u}rst}, F., {Barret}, D., {Beloborodov}, A., {Boggs},
  S.~E., {Christensen}, F.~E., {Craig}, W.~W., {Fabian}, A.~C., {Hailey},
  C.~J., {Hornschemeier}, A., {Kaspi}, V., {Kulkarni}, S.~R., {Maccarone}, T.,
  {Miller}, J.~M., {Rana}, V., {Stern}, D., {Tendulkar}, S.~P., {Tomsick}, J.,
  {Webb}, N.~A., \& {Zhang}, W.~W. 2014, \nat, 514, 202

\bibitem[{{Bardou} \& {Heyvaerts}(1996)}]{1996A&A...307.1009B}
{Bardou}, A. \& {Heyvaerts}, J. 1996, \aap, 307, 1009

\bibitem[{{Basko} \& {Sunyaev}(1976)}]{1976MNRAS.175..395B}
{Basko}, M.~M. \& {Sunyaev}, R.~A. 1976, \mnras, 175, 395

\bibitem[{{Basko} \& {Syunyaev}(1975)}]{1975JETP...41...52B}
{Basko}, M.~M. \& {Syunyaev}, R.~A. 1975, Soviet Journal of Experimental and
  Theoretical Physics, 41, 52

\bibitem[{{Braun} \& {Yahel}(1984)}]{1984ApJ...278..349B}
{Braun}, A. \& {Yahel}, R.~Z. 1984, \apj, 278, 349

\bibitem[{{Campbell}(1987)}]{1987MNRAS.229..405C}
{Campbell}, C.~G. 1987, \mnras, 229, 405

\bibitem[{{Dotan} \& {Shaviv}(2011)}]{2011MNRAS.413.1623D}
{Dotan}, C. \& {Shaviv}, N.~J. 2011, \mnras, 413, 1623

\bibitem[{{Fabbiano} \& {White}(2006)}]{2006csxs.book..475F}
{Fabbiano}, G. \& {White}, N.~E. {Compact stellar X-ray sources in normal
  galaxies}, ed. W.~H.~G. {Lewin} \& M.~{van der Klis}, 475--506

\bibitem[{{Feng} \& {Soria}(2011)}]{2011NewAR..55..166F}
{Feng}, H. \& {Soria}, R. 2011, \nar, 55, 166

\bibitem[{{Frank} {et~al.}(2002){Frank}, {King}, \&
  {Raine}}]{2002apa..book.....F}
{Frank}, J., {King}, A., \& {Raine}, D.~J. 2002, {Accretion Power in
  Astrophysics: Third Edition}

\bibitem[{{Ghosh} \& {Lamb}(1979{\natexlab{a}})}]{1979ApJ...232..259G}
{Ghosh}, P. \& {Lamb}, F.~K. 1979{\natexlab{a}}, \apj, 232, 259

\bibitem[{{Ghosh} \& {Lamb}(1979{\natexlab{b}})}]{1979ApJ...234..296G}
---. 1979{\natexlab{b}}, \apj, 234, 296

\bibitem[{{Harding} {et~al.}(1984){Harding}, {Kirk}, {Galloway}, \&
  {Meszaros}}]{1984ApJ...278..369H}
{Harding}, A.~K., {Kirk}, J.~G., {Galloway}, D.~J., \& {Meszaros}, P. 1984,
  \apj, 278, 369

\bibitem[{{Kaaret} {et~al.}(2003){Kaaret}, {Corbel}, {Prestwich}, \&
  {Zezas}}]{2003Sci...299..365K}
{Kaaret}, P., {Corbel}, S., {Prestwich}, A.~H., \& {Zezas}, A. 2003, Science,
  299, 365

\bibitem[{{Karino} {et~al.}(2008){Karino}, {Kino}, \&
  {Miller}}]{2008PThPh.119..739K}
{Karino}, S., {Kino}, M., \& {Miller}, J.~C. 2008, Progress of Theoretical
  Physics, 119, 739

\bibitem[{{Klu{\'z}niak} \& {Rappaport}(2007)}]{2007ApJ...671.1990K}
{Klu{\'z}niak}, W. \& {Rappaport}, S. 2007, \apj, 671, 1990

\bibitem[{{Koldoba} {et~al.}(2002){Koldoba}, {Lovelace}, {Ustyugova}, \&
  {Romanova}}]{2002AJ....123.2019K}
{Koldoba}, A.~V., {Lovelace}, R.~V.~E., {Ustyugova}, G.~V., \& {Romanova},
  M.~M. 2002, \aj, 123, 2019

\bibitem[{{Lamb} {et~al.}(1973){Lamb}, {Pethick}, \&
  {Pines}}]{1973ApJ...184..271L}
{Lamb}, F.~K., {Pethick}, C.~J., \& {Pines}, D. 1973, \apj, 184, 271

\bibitem[{{Mapelli} {et~al.}(2010){Mapelli}, {Ripamonti}, {Zampieri}, {Colpi},
  \& {Bressan}}]{2010MNRAS.408..234M}
{Mapelli}, M., {Ripamonti}, E., {Zampieri}, L., {Colpi}, M., \& {Bressan}, A.
  2010, \mnras, 408, 234

\bibitem[{{Medvedev} \& {Poutanen}(2013)}]{2013MNRAS.431.2690M}
{Medvedev}, A.~S. \& {Poutanen}, J. 2013, \mnras, 431, 2690

\bibitem[{{Naso} {et~al.}(2013){Naso}, {Klu{\'z}niak}, \&
  {Miller}}]{2013MNRAS.435.2633N}
{Naso}, L., {Klu{\'z}niak}, W., \& {Miller}, J.~C. 2013, \mnras, 435, 2633

\bibitem[{{Pringle} \& {Rees}(1972)}]{1972A&A....21....1P}
{Pringle}, J.~E. \& {Rees}, M.~J. 1972, \aap, 21, 1

\bibitem[{{Rappaport} {et~al.}(2004){Rappaport}, {Fregeau}, \&
  {Spruit}}]{2004ApJ...606..436R}
{Rappaport}, S.~A., {Fregeau}, J.~M., \& {Spruit}, H. 2004, \apj, 606, 436

\bibitem[{{Roberts}(2007)}]{2007Ap&SS.311..203R}
{Roberts}, T.~P. 2007, \apss, 311, 203

\bibitem[{{Romanova} {et~al.}(2002){Romanova}, {Ustyugova}, {Koldoba}, \&
  {Lovelace}}]{2002ApJ...578..420R}
{Romanova}, M.~M., {Ustyugova}, G.~V., {Koldoba}, A.~V., \& {Lovelace},
  R.~V.~E. 2002, \apj, 578, 420

\bibitem[{{Shakura} \& {Sunyaev}(1973)}]{ShakuraSunyaev}
{Shakura}, N.~I. \& {Sunyaev}, R.~A. 1973, \aap, 24, 337

\bibitem[{{Shaviv}(1998)}]{1998ApJ...494L.193S}
{Shaviv}, N.~J. 1998, \apjl, 494, L193

\bibitem[{{Sutton} {et~al.}(2013){Sutton}, {Roberts}, \&
  {Middleton}}]{2013MNRAS.435.1758S}
{Sutton}, A.~D., {Roberts}, T.~P., \& {Middleton}, M.~J. 2013, \mnras, 435,
  1758

\bibitem[{{Sutton} {et~al.}(2012){Sutton}, {Roberts}, {Walton}, {Gladstone}, \&
  {Scott}}]{2012MNRAS.423.1154S}
{Sutton}, A.~D., {Roberts}, T.~P., {Walton}, D.~J., {Gladstone}, J.~C., \&
  {Scott}, A.~E. 2012, \mnras, 423, 1154

\bibitem[{{Thompson} \& {Duncan}(1995)}]{TD95}
{Thompson}, C. \& {Duncan}, R.~C. 1995, \mnras, 275, 255

\bibitem[{{Wang}(1987)}]{1987A&A...183..257W}
{Wang}, Y.-M. 1987, \aap, 183, 257

\end{thebibliography}

\end {document}